\documentclass[a4paper,superscriptaddress, 12pt,eqsecnum,floats,aps,amsmath,amssymb,nofootinbib,prd]{article}
\usepackage{geometry}
\usepackage{authblk}
\usepackage{setspace}
\usepackage{amsmath,amssymb,amsfonts,amsthm,mathrsfs}
\usepackage{graphicx}
\usepackage{enumerate} 
\usepackage{comment}
\usepackage[pdftex]{hyperref}

\def\be{\begin{equation}}
\def\ee{\end{equation}}
\def\ba{\begin{eqnarray}}
\def\ea{\end{eqnarray}}
\def\bi{\begin{itemize}}
\def\ei{\end{itemize}}

\def\reals{\mathbb{R}}

\def\bra{\langle}
\def\ket{\rangle}

\def\O{\Omega}
\def\s{\sigma}
\def\sign{{\rm sign}}
\def\L{\mathcal{L}}
\def\w{\omega}
\def\xh{\hat{x}}
\def\zb{\bar{z}}
\def\deltat{\delta^{(2)}}

\def\kh{\hat{k}}

\def\khi{\kh^{\rm in}}
\def\kho{\kh^{\rm out}}
\def\Ein{E^{\rm in}}
\def\Eout{E^{\rm out}}

\def\Ih{\hat{\mathcal{I}}}
\def\qh{\hat{q}}

\def\I{\mathcal{I}}

\def\Do{\mathring{D}}
\def\Hout{H^{\rm out}}
\def\Hin{H^{\rm in}}

\def\Fout{\mathcal{H}^{\rm out}}
\def\Fin{\mathcal{H}^{\rm in}}

\def\O{\Omega}
\def\L{\mathcal{L}}
\def\I{\mathcal{I}}
\def\e{\epsilon}
\def\tf{\text{TF}}
\def\xio{\mathring{\xi}}

\def\Xhom{X^{\rm hard}}
\def\Xinhom{X^{\rm soft}}
\def\Hhom{H^{\rm hard}}
\def\Hinhom{H^{\rm soft}}
\def\lhs{\text{LHS}}
\def\eh{\hat{\epsilon}}

\def\G{\mathbf{G}}

\title{Asymptotic symmetries and subleading soft graviton theorem}
\author[a]{Miguel Campiglia}
\author[b]{Alok Laddha}
\affil[a]{Raman Research Institute}
\affil[b]{Chennai Mathematical Institute}
\date{}
\begin{document}
\maketitle

\thispagestyle{empty}

\let\oldthefootnote\thefootnote
\renewcommand{\thefootnote}{\fnsymbol{footnote}}
\footnotetext{Email: miguel@rri.res.in, aladdha@cmi.ac.in}
\let\thefootnote\oldthefootnote

\begin{abstract}
Motivated by the equivalence between soft graviton theorem and Ward identities
for the supertranslation symmetries belonging to the BMS group, we propose a new
extension (different from the so-called extended BMS) of the BMS group which is a
semi-direct product of supertranslations  and ${\rm Diff}(S^2)$. We propose a definition for the canonical generators associated
to the smooth diffeomorphisms and show that the resulting Ward identities are
equivalent to the subleading soft graviton theorem of Cachazo and Strominger.
\end{abstract}
%%%%%%%%%%%%%%%%%%%%%%%%%%%%%%%%%%%%%%%%%%%%%%%%%%%%%%%%%%%%%%
\maketitle

\section{Introduction}
It has been  known since 60s that there is an infinite dimensional symmetry group underlying asymptotically flat spacetimes  known as the BMS group \cite{bms, sachs}.
The role of  BMS group in  quantum theory was elucidated in a  series of remarkable papers by Ashtekar et. al. \cite{aaprl,AS,aajmp}. In \cite{aaprl} the radiative modes of the full non-linear gravitational field were isolated and equipped with a symplectic structure, thus paving the way for (asymptotic) quantization of gravity. In \cite{AS}, it was shown that the BMS group is a dynamical symmetry group of the radiative phase space and the corresponding Hamiltonians were obtained. The reasons behind the enlargement of the translation subgroup (of the Poincare group) to supertranslations was clarified in \cite{aajmp}, where it was shown that the space of `vacuum configurations' (i.e. points in phase space for which the fluxes of all BMS momenta vanish identically) are in one to one correspondence with supertranslations (modulo translations). This in turn led to the first detailed relation between the BMS supertranslations and the infrared issues in quantum gravity  \cite{aaoxf,aabook}. In particular, it clarified the need to use coherent states which lead to an S-matrix free of infrared divergences \cite{kf,akhoury}.

%in order to obtain an infrared finite  S-matrix \cite{kf,akhoury}.

In recent months there has been a renewed interest in analyzing these symmetries in the context of quantum gravity S-matrix. 
There are two reasons for this resurgence. First being a series of fascinating papers by Strominger et. al. \cite{strominger1,strominger2,st} where a precise relationship between  Ward identities associated with  supertranslation symmetries and  Weinberg's soft graviton theorem \cite{weinberg} was unravelled. %\footnote{Note that the infrared-BMS relationship explored in these works is a priori different  from the one described in the previous paragraph. In particular,  whereas Ashtekar's approach is related to  coherent states yielding IR-finite S-matrix \cite{kf,akhoury},   Strominger's  approach characterizes the IR-divergences of the standard Fock states  S-matrix.} 
The second reason is an extremely interesting proposal by Barnich and Troessaert \cite{barnich1,barnich2,barnich3} that this symmetry can be naturally extended to include the Virasoro group, which in turn may shed new light on duality between quantum gravity in the bulk and conformal field theory on the boundary. In the literature this group is referred to as the extended BMS group.

The two ideas mentioned above converged in \cite{virasoro} where it was shown that the Ward identities associated to precisely such Virasoro symmetries follow from the so-called subleading soft theorem for gravitons. This theorem, conjectured by Strominger, was proved at  tree-level in the so-called holomorphic soft limit in \cite{cs}, where its validity was also checked in a number of examples. A more general proof for the theorem was later given in \cite{plefka,bern}. See \cite{gj,naculich,white,sterman,beneke} for earlier works on soft graviton amplitudes  and \cite{bal1,bal2,schwab,afkami,zlotnikov,rojas,skinner,mason} for an incomplete list of recent related works.

However as noted in \cite{virasoro}, whereas for the  supertranslation symmetries the Ward identities are in fact equivalent to  Weinberg's soft graviton theorem, such an equivalence could not be established as far as the Virasoro symmetries and the subleading theorem were concerned. Motivated by the need to establish such an equivalence, in this paper we propose a different extension of the BMS group. Instead of extending the global conformal symmetries to the Virasoro symmetries as in \cite{barnich1}, we extend them to smooth vector fields on the sphere. We refer to this group as the generalized BMS group and denote it by $\G$. We show  $\G$ is the semi-direct product of supertranslations with smooth diffeomorphisms of the conformal sphere ($\textrm{Diff}(S^{2})$) and that it preserves the space of asymptotically flat solutions to Einstein's equations. However, contrary to the BMS group it does not preserve the leading order \emph{kinematical} metric components, for instance  by generating arbitrary diffeomorphisms of the conformal sphere at infinity. We define  charges associated to this symmetry ($\textrm{Diff}(S^{2})$) in the radiative phase space of the gravitational field. Our definition of these charges is motivated by the charges one obtains for extended BMS symmetry. Although this definition is ad-hoc and not derived by systematic analysis, 
 we show  its associated  ``Ward identities"  are in one-to-one correspondence with the subleading soft graviton theorem. The analysis performed here is rather similar in spirit to the recent work by Lysov, Pasterski and Strominger for  massless QED  \cite{lowsym}.  Exactly as in that case, our charges do not form a closed algebra. We leave the interpretation of this non-closure  for future investigations.

The outline of this paper is as follows. In section \ref{sec2} we define $\G$ and show that it preserves asymptotic flatness.  We show $\G$ can be characterized as the group of diffeomorphism that preserve null infinity and which are asymptotically volume preserving. In section \ref{sec3} we review the radiative phase space formulation of Ashtekar  and show how the action of extended BMS is Hamiltonian\footnote{Modulo certain subtleties related to the IR sector.}. We emphasize on using the radiative phase space framework carefully since, as illustrated in appendix \ref{PBapp}, the weakly non-degeneracy of the symplectic structure implies that certain seemingly natural Poisson bracket relations are ill-defined and their use can lead to incorrect results.\\
Just as the BMS group can be defined purely in terms of structures available at null infinity without referring to  spacetime, we present $\G$ from the perspective of null infinity in section \ref{sec4}. In this section we also present our prescription for the Hamiltonian action of the generators of $\G$ on the radiative phase space of gravity.  In section \ref{sec5} we analyze the ``Ward identities''   associated to this prescription and show their  equivalence with the subleading soft graviton theorem.

\section{Spacetime picture} \label{sec2}
\subsection{Proposal for a generalization of the BMS group}

Let us for concreteness focus on future null infinity ${\cal I}^{+}$. Following \cite{strominger2} we refer to the algebra of asymptotic symmetries at ${\cal I}^{+}$ as $\textrm{BMS}^{+}$. In the original  derivation of BMS algebra, through an interplay between fall-off conditions and Einstein equations, one arrives at  the following form of asymptotically flat metrics (we take expressions from  \cite{strominger2,barnich2}):
\begin{multline}
ds^2 = (1 + O(r^{-1})) du^2 - (2+ O(r^{-2}))d u d r + \\ (r^2 q_{AB} +r C_{AB}(u,\xh) +O(1)) dx^A d x^B +O(1) dx^A du  .\label{bmsfalloff}
\end{multline}
Here  $x^A$ are coordinates on the 2-sphere, $q_{AB}$ the round unit sphere metric (whose covariant derivative we denote by $D_A$) and $\xh$ denotes points on the sphere. $C_{AB}$ is trace-free and unconstrained by Einstein equations, whereas the remaining  metric components are determined by $C_{AB}$ through Einstein's equations. $C_{AB}$ is referred to as the free gravitational radiative data.

$\textrm{BMS}^{+}$ is defined as the algebra of vector fields which preserve the fall-offs (\ref{bmsfalloff}).  It is generated by vector fields of the (asymptotic) form,
\be
\xi_+^{a}\partial_{a}\ =\ V_+^{A} \partial_A + u \alpha_{+}  \partial_u  - r \alpha_{+} \partial_r + f_{+} \partial_u + \lambda_+^a \partial_a  , \label{bmsxi}
\ee
where  $V_+^{A}$ is a conformal Killing vector field (CKV) of the sphere, $\alpha_{+} = (D_A V_+^{A})/2$ and $f_{+}=f_{+}(\xh)$ any smooth function on the sphere. $\lambda_+^a \partial_a$ denotes the next terms in the  $1/r$ expansion \cite{barnich2}:
\be
 \lambda_+^a \partial_a = -\frac{u}{r} D^A\alpha \, \partial_A +\frac{u}{2} D_B D^B \alpha \, \partial_r + \ldots
\ee
One can similarly define the algebra $\textrm{BMS}^{-}$ of asymptotic symmetries associated to past null infinity ${\cal I}^{-}$.

In \cite{strominger2} Strominger introduced the remarkable notion of $\textrm{BMS}^{0} \subset \textrm{BMS}^{+}\times\textrm{BMS}^{-}$, which he argued to be a symmetry of quantum gravity S-matrix. This group maps an incoming scattering data, characterized by fields on ${\cal I}^{-}$, to outgoing scattering data, characterized by fields on ${\cal I}^{+}$, while conserving total energy. Identifying the null generators of ${\cal I}^{+}$ and ${\cal I}^{-}$ through ${\cal I}^{+}|_{u=-\infty}\ =\ {\cal I}^{-}|_{v=+ \infty}\ = i^{0}$, the group is defined  by the conditions  \cite{strominger2}:
\be 
V_+^{A}(\xh)\ =\ V_-^{A}(\xh), \quad f_{+}(\xh)\ =\ f_{-}(\xh) \label{bmsnot}.
\ee
We now consider the scenario where in $\xi_+^a$ given in (\ref{bmsxi}),  $V_+^{A}$ is not CKV. A simple computation reveals that under the diffeomorphisms generated by such vector fields, the metric coefficients whose fall-offs  are violated are,
\begin{equation}
\begin{array}{lll}
{\cal L}_{\xi^{+}}g_{AB}\ =\ O(r^{2})\\
\vspace*{0.1in}
{\cal L}_{\xi_{+}}g_{u u} =  O(1).
\end{array}
\end{equation}
Thus, relaxing the CKV condition forces us to consider metrics  where the $O(r^{2})$ part of $g_{AB}$ is not  necessarily the round metric and such that  $\ g_{uu}\ =\ O(1)$.
 We are thus lead to consider metrics of the form:\footnote{The form (\ref{gralfalloff}) is of the type of metrics  considered in  \cite{barnich2} except that we take $q_{AB}$ to be  $u$-independent and we do not require $q_{AB}$ to be a conformal rescaling of the unit round metric.}
\be
ds^2 = O(1) du^2 - (2+O(r^{-2}) )d u d r + (r^2 q_{AB} +O(r)) dx^A d x^B +O(1) dx^A du , \label{gralfalloff}
\ee
with $q_{AB}$  no longer the standard metric on $S^{2}$. 
%In particular, $r$ does not  necessarily coincide with the $r_0 \to \infty$ areal radius of the two-spheres $u=u_0$, $r=r_0$.
We can now ask if these spacetimes with more general fall-offs of the metric coefficients are asymptotically flat. As shown in \cite{aabook} the answer is in the  affirmative.  This can be most easily seen from the conformal description of asymptotic flatness. In this description, asymptotic flatness is captured by the existence of a conformal factor $\O$ such that $\O^2 ds^2$ has a well defined limit at null infinity and satisfies a number of properties. It can be shown  that such spacetimes admit coordinates in a neighborhood of null infinity for which the metric fall-offs include those of the form (\ref{gralfalloff}), with $\O \sim 1/r $ \cite{aabook,aaunpublished}.
%by the fact that under conformal completion with $\omega\ =\ \frac{1}{r}$ this leads to a a unphysical metric which satisfies all the criteria of asymptotic flatness (namely a existence of complete null infinity with topology $S^{2}\times{\bf R}$ \cite{aabook})

We refer to this group of asymptotic symmetries at future null infinity as  generalized $\textrm{BMS}^{+}$ group and denote it by $\G^{+}$.  $\G^{+}$ is a semi-direct product of supertranslations and $\textrm{Diff}(S^{2})$, with supertranslations being a normal Abelian subgroup exactly as in the case of the BMS group.
 
 One can similarly define a corresponding group associated to ${\cal I}^{-}$ and we refer to it as $\G^{-}$. Following the strategy used for the BMS \cite{strominger2} and extended BMS \cite{virasoro} cases,  we define the subgroup $\G^0$ of  $\G^{+}\times\G^{-}$ by the  identification (\ref{bmsnot}) for generators of $\G^{+}$ and $\G^{-}$.
 It then follows that  $\G^0$ reduces to $\textrm{BMS}^{0}$ when $V^A$ is  CKV. 
 % and\\ \noindent{(2)}  $\lim_{u\rightarrow\ -\infty}{\cal L}_{\xi^{+}}g_{uu}\ =\ lim_{v\rightarrow\ \infty}{\cal L}_{\xi^{-}}g_{vv}$.\\ The second condition is a generalization of the continuity of the Bondi-mass aspect across ${\cal I}^{+}|_{u=-\infty}\ =\ {\cal I}^{-}|_{v=+ \infty}\ = i^{0}$.\\
%Using constraints equation which relate News tensor and matter stress tensor to Bondi mass aspect, it follows that second condition implies that the total incoming energy equals total outgoing energy. Hence this group can be associated to (low energy) scattering processes.\\

From now on we drop the labels $+,-$  and parametrize the generalized BMS vector fields by $(V^{A}, f)$.

\subsection{Characterization of $\G$}
We now  ask if there is any geometrical characterization of the generalized BMS vector fields. Recall that BMS vector fields can be characterized as asymptotic Killing vector fields: 
\be
\nabla_{(a} \xi_{b)}  \to 0 \quad \text{as} \quad r \to \infty.
\ee
Whereas generalized BMS clearly do not satisfy this condition, it turns out they are asymptotic divergence-free vector fields:
\be
\nabla_a \xi^a  \to 0 \quad \text{as} \quad r \to \infty.   \label{asymvolzero}
\ee
Indeed, a simple calculation shows:
\ba
\nabla_a \xi^a & = &  \partial_{u}\xi^{u}\ +\ D_{A}V^{A}\ +\ \frac{1}{r^{2}}\partial_{r}(r^{2}\xi^{r})\ +\ O(r^{-1}) \\
& = & \alpha\ +\ 2\alpha\ -\ 3\alpha + O(r^{-1})\ \\
& = & \ O(r^{-1}).
\ea
We now show the converse, namely that generalized BMS vector fields are characterized by (\ref{asymvolzero}) and the preservation of the fall-offs (\ref{gralfalloff}). A general vector field preserving  $\I$ has the following general form as $r \to \infty$:
\be
\xi^a = \xio^A(u,\xh) \partial_A + \xio^u(u,\xh) \partial_u + r \xio^r(u,\xh) \partial_r + \ldots  \label{gralxi}
\ee
where the dots indicate terms of the form: $O(r^{-1}) \partial_A + O(r^{-1}) \partial_u + O(1) \partial_r$.   We only focus on the leading terms in the $1/r$ expansion.  Subleading terms are determined  by requiring  preservation of the fall-offs (\ref{gralfalloff}), and their  form  depend on the specific metric coefficients in (\ref{gralfalloff}).  \\
Eq. (\ref{asymvolzero}) gives:
\be
\nabla_a \xi^a  = O(r^{-1}) \iff D_A \xio^A + \partial_u \xio^u + 3 \xio^r =0\label{volzero} .
\ee
The components of (\ref{gralfalloff}) leading to restrictions on the leading part of  (\ref{gralxi}) are:
\be
\L_{\xi}g_{ur}= O(r^{-1}) \iff  \partial_u \xio^u + \partial_r \xio^r =0 \label{guu}
\ee
\be
\L_{\xi}g_{uA}= O(r), \;  \L_{\xi}g_{uu}= O(1) \iff  \partial_u \xio^A=0,  \;  \partial_u \xio^r=0 .\label{uindep}
\ee
It is easy to verify that the most general solution to Eqns. (\ref{volzero}), (\ref{guu}), (\ref{uindep}) is given by:
\ba
\xio^A(u,\xh)   & = & V^A(\xh) \\
\xio^u(u,\xh) & = & u \alpha(\xh) + f(\xh) \\
\xio^r(u,\xh) & = & - \alpha(\xh)
\ea
with $V^A(\xh)$ and $f(\xh)$ undetermined and $ \alpha = (D_A V^A)/2$ as before.  Thus, we recover the leading term of (\ref{bmsxi}) with the CKV condition on $V^A$ dropped. This precisely represents the proposed  generalized BMS vector fields.  The preservation of (\ref{gralxi}) for the remaining metric components impose conditions on the subleading terms of $\xi^a$ indicated by the dots in (\ref{gralxi}).

\subsection{Difficulties in extracting a map on radiative data}\label{sec2.3}
We recall that BMS vector fields have a well defined action on the unconstrained radiative data characterized by $C_{AB}$. For $\xi^a$ as in (\ref{bmsxi}) with $f=0$ the action is given by \cite{strominger2}:
\begin{equation}
\delta_V C_{AB}= \L_V C_{AB} - \alpha C_{AB} + \alpha u \partial_u C_{AB} .
\end{equation}
Although  generalized BMS vector fields map an asymptotically flat spacetime to another one, they do not induce any obvious map on the free radiative data. 
 As they change the zeroth order structure, the linear in $r$ coefficients of $g_{AB}$ do not represent all free data. 

In order to bring out the differences with the BMS case,  consider the action of generalized BMS vector field on the  $g_{AB}$ metric components (again we consider the case $f=0$ and $g_{ab}$ as in (\ref{bmsfalloff})):
\be\label{deltavcab}
\L_{\xi}g_{AB} = r^2 ( \L_V q_{AB} - 2 \alpha q_{AB}) + 
r ( \L_V C_{AB} - \alpha C_{AB} + \alpha u \partial_u C_{AB}) + u r s_{AB} ,
\ee
where
\be\label{softjuly28}
s_{AB}= -2 D_A D_B \alpha + D_C D^C \alpha \, q_{AB}.
\ee
Since the zeroth order structure changes, the action of generalized BMS encodes the physical transformations (i.e. change in the radiative data) as well as ``gauge transformations" induced by the change in the zeroth order structure. It is not clear to us how to extract out the gauge-invariant change in the News from this action. This point will be important in defining the action of generalized BMS operator in quantum theory. We return to this issue in section \ref{sec4C}.

\section{Radiative phase space}\label{sec3}
\subsection{Review of Ashtekar formulation} \label{sec3.1}
In this section we recall Ashtekar's description of the radiative phase space of gravity following mostly references \cite{AS,aabook}.  We will only present the end result of the description, and   encourage the reader to look at  \cite{aaprl,AS,aabook} for its motivation from spacetime perspective, as well as reference \cite{am} for its relation with  ADM phase space.

The idea is to start with $\I$ (which will eventually stand for either future or past null infinity) as an abstract 3-manifold, topologically $S^2 \times \reals$, and ruled by preferred directions or `rays' so that there is a canonical projection $\I \to  \Ih \sim S^2$ with $\Ih$ the space of rays.  Next, one endows $\I$  with a `universal structure' which plays the role of kinematical arena. This universal structure is given by an equivalence class of pairs $(q_{ab},n^a)$ where $n^a$ is a vector field tangent to the rays and $q_{ab}$  a $(0,+,+)$ degenerate metric that is given by the pull-back of a 2-metric $\qh_{ab}$ on $\Ih$,  so that $q_{ab}n^b=0$ and $\L_n q_{ab}$=0. Each pair is referred to as a `frame'. The equivalence is given by:
\be
(q_{ab},n^a) \sim  (\w^2 q_{ab},\w^{-1} n^a) ,\quad \forall \w: \I \to \reals : \L_n \w =0, \label{equivframes}
\ee
and the corresponding equivalence class  $[(q_{ab},n^a)]$ gives the `universal structure'. The BMS group discussed in the previous section arises in this context as the group of diffeomorphism of $\I$ that preserve this universal structure \cite{aabook}. 

We now describe the dynamical degrees of freedom and associated phase space.  The description uses a fixed `frame' $(q_{ab},n^a) \in [(q_{ab},n^a)]$, so that, strictly speaking, one arrives at a family of phase spaces parametrized  by the frames $(q_{ab},n^a) \in [(q_{ab},n^a)]$. One then shows that there exists a natural isomorphism between the different phase spaces associated to the different frames.  Below we present the phase space associated to a given frame. The isomorphism, crucial for the implementation of boosts in phase space, is described in appendix \ref{boostapp}.

A derivative operator $D_a$ on $\I$ is said to be compatible with a frame  $(q_{ab},n^a)$ if it satisfies:
\be
D_c q_{ab}=0, \quad D_a n^b =0, \quad  2 D_{(a} V_{b)}= \L_{V} q_{ab}  \;  \; {\rm if } \; \; V_a n^a=0, \label{compatibleD}
\ee
where the Lie derivative is along any vector $V^a$ satisfying $V_a= q_{ab}V^b$.  Introduce the following equivalence relation on derivative operators satisfying (\ref{compatibleD}):\footnote{This equivalence relation is unrelated to the one in (\ref{equivframes}). From the spacetime perspective, (\ref{equivframes}) arises from different values the conformal factor $\O$ can take at $\I$, whereas (\ref{compatibleD}) arises from different values the derivative of the  conformal factor (along directions off $\I$) can take at $\I$.}
\be
D'_a  \sim D_a  \quad {\rm if } \quad D'_a k_b = D_a k_b + f n^c k_c q_{ab} \label{equivalenceD}
\ee
for some function $f$.  The  phase space, denoted by $\Gamma$, is the space of equivalence classes $[D_a]$ of (torsion-free) derivative operators satisfying (\ref{compatibleD}).  A parametrization of this space is obtained as follows. Fix a derivative $\Do_a$ satisfying (\ref{compatibleD}). It can be shown that any other derivative $D_a$  satisfying (\ref{compatibleD}) is given by:
\be
D_a k_b =\Do_a k_b+  (\Sigma_{ab} n^c)k_c ,\label{defSigma}
\ee
where $\Sigma_{ab}$ is symmetric and satisfies $\Sigma_{ab} n^b=0$.  Such tensors parametrize the space of connections $D_a$ compatible with $(q_{ab},n^a)$. From (\ref{equivalenceD}) it follows that 
\be
\sigma_{ab} := \Sigma_{ab}-\frac{1}{2} q_{ab} q^{cd}\Sigma_{cd} = ((D_a - \Do_a) k_b )^\tf  \quad \text{for any } k_b \, :  \, n^b k_b = 1  , \label{defsigma}
\ee
can be used to parametrize the space of equivalence classes $[D_a]$.  We recall that $q^{ab}$ is defined up to $v^{(a}n^{b)}$ so that the trace-free symbol `TF' is only  well defined on tensors annihilated by $n^a$.
% such equivalence classes can be  given by symmetric tensors $\sigma_{ab}$ satisfying $\sigma_{ab} n^b=0$ and $\sigma_{ab} q^{ab}=0$, where $\sigma_{ab}$ represeting the trace free part of $\Sigma_{ab}$  defined in  (\ref{defSigma}):
In terms of this parametrization  the symplectic structure reads:
\be
\O(\sigma_1 , \sigma_2) = \int d^3 V q^{ac} q^{bd} ( \s^1_{ab}  \L_n \s^2_{cd} - \s^2_{ab} \L_n \s^1_{cd})  \label{sympstr},
\ee
where $d^3V = \e_{abc}$, with $\e_{ab} =\e_{abc}n^c$ the area form of $q_{ab}$. 

Let us now make contact with the spacetime picture of section \ref{sec2}. For concreteness we focus in future null infinity. For spacetime metrics as in (\ref{bmsfalloff}),  $\I$ is described by the coordinates $(u,x^A)$ with  $n^a\partial_a = \partial_u$ and $q_{ab} d x^a d x^b= q_{AB} d x^A d x^B$. One can verify that the nonzero components of $\sigma_{ab}$ are: $\sigma_{AB}= (1/2) C_{AB}$.  The News tensor is then given by\footnote{Our convention for the News tensor, taken from \cite{AS}, differs by a sign from that used in \cite{st,virasoro}.}
\be
N_{AB}(u,\xh)= -2 \dot{\sigma}_{AB}(u,\xh), \label{news}
\ee
where $\dot{\sigma}_{AB} \equiv \L_n \sigma_{AB} \equiv \partial_u \sigma_{AB}$. 

%In appendix \ref{modefnapp} we recover the fact that the Poisson brackets arising from (\ref{sympstr}) match with those arising from the perspective of perturbative gravity. %In subsection \ref{pbssec}  and appendix \ref{PBapp} we discussion of `zero mode' subtleties with the Poisson brackets).

We conclude by describing the fall-offs of radiative phase space. In $(u,x^A)$  coordinates 
%The fall-off conditions we shall need for our purposes are as follows.   In the $(u,x^A)$ be coordinates on $\I$ as before  @@@@see where these are first introduce@@@@, phase space is given by   smooth tensors $\sigma_{AB}(u,\xh)$  satisfying the algebraic conditions above ($\sigma_{[AB]}= \sigma_{AB} q^{AB}=0$) with the following fall-off at infinity:
they are  given by  \cite{AS}:
\be
\sigma_{AB}(u,\xh) = \sigma^{\pm}_{AB}(\xh) + O(u^{-\e}) \quad \text{as} \quad u \to \pm \infty, \label{falloffs}
\ee
where $\e>0$ and the limiting values  $\sigma^{\pm}_{AB}(\xh)$ are kept unspecified (but smooth in $\xh$). These fall-offs  ensure  the convergence of the integral defining the symplectic structure (\ref{sympstr}).\footnote{The fall-offs used  by Strominger based on the analysis of CK spaces corresponds to $\e=1/2$ \cite{strominger2}. It thus seems that the range   $0 < \e < 1/2$  is not relevant for gravitational scattering. We nevertheless keep $\e$ general as all we need in our analysis is $\e>0$.}

\subsubsection{Poisson brackets subtleties} \label{pbssec}
We  comment on a subtlety associated to the Poisson brackets that was noticed in \cite{st}.   From the radiative phase space perspective the  symplectic form (\ref{sympstr}) is the fundamental structure whereas Poisson brackets are  derived quantities. We recall that in this approach the Hamiltonian vector field (HVF) $X_F$ of a phase space function $F$ is defined as the solution to the equation 
\be
\O( X_F, \cdot ) = d F, \label{defhvf}
\ee
 and that,  given two phase space functions $F$ and $G$ admitting HVFs, their Poisson bracket is defined by $\{ F, G \} := \O(X_F,X_G) = X_G(F)$.  In \cite{AS} it is shown that $\O$ is weakly non-degenerate, that is,  $\O$ considered as a map from $T \Gamma$ to $T^* \Gamma$ is injective but not necessarily  surjective. Thus, there is no guarantee that one can always solve Eq. (\ref{defhvf}) (but if there is a solution, it is unique).  As discussed in appendix \ref{PBapp}, an example of a function not admitting a HVF is given by $F[\sigma]:= \int_{\I} d^3 V F^{AB}(u,\xh) \sigma_{AB}(u,\xh)$ with  $\int_{-\infty}^\infty du F^{AB}(u,\xh) \neq 0$.  In particular, one cannot define PBs between $\sigma_{AB}(u,\xh) $ and $\sigma_{AB}(u',\xh')$. Fortunately, these `undefined PBs' are nowhere needed in the analysis.

\subsection{(Extended) BMS action on $\Gamma$} \label{sec4B}
Let $D_a$ be a connection as in (\ref{compatibleD}) with $[D_a]$ the corresponding element in  radiative phase space.  Under the action of  a BMS vector field  $\xi^a$ the connection changes by $\delta_\xi D_a= [\L_\xi,D_a]$. If $\xi^a$ preserves the frame (case of supertranslations and rotations), the transformed connection $D'_a \approx D_a +\delta_\xi D_a$ is compatible with the frame and one can directly read off the phase space action from $\delta_\xi D_a$. For boosts however, the transformed connection is compatible with the frame $(q'_{ab}, n'^a) \approx ((1+ 2\alpha) q_{ab}, (1-\alpha) n^a)$. One thus needs to use the isomorphism between the phase spaces associated to the  different frames in order to obtain the phase space action. The resulting action reads (see appendix \ref{boostapp} for its derivation):
\begin{equation}\label{boost1}
\begin{array}{lll}
(X_{\xi})_{ab} =  ([\L_\xi,D_a]k_{b} +2 k_{(a} \partial_{b)} \alpha)^{\tf},
\end{array}
\end{equation}
where $k_{a}$ is any covector satisfying $n^{a} k_{a}\ =\ 1$. %One can verify that the `correction term' $2 k_{(a} \partial_{b)$ ensures that the total tensor in brackets is annihilated by $n^a$ so that `tracein %For computational purposes we often choose $k_{a}\ =\ l_{a}$.\\

In $(u,x^A)$ coordinates, for a `pure rotation/boost' vector field 
\be
\xi^a \partial_a = V^A \partial_A + u \alpha \partial_u, \label{boostrot}
\ee  
the expression takes the form:
\be
(X_{V})_{AB} = \L_V \sigma_{AB} - \alpha \sigma_{AB} +u \alpha \dot{\sigma}_{AB} - u (D_A D_B \alpha )^{\tf} .\label{XxiAB} 
\ee
Following \cite{st,virasoro}, we refer to the piece linear in $\sigma$  as the `hard term' and the $\sigma$-independent, linear in $u$ piece as the `soft term'.  The soft term appears to violate the fall-offs (\ref{falloffs}). However the CKV nature of $V^{A}$ implies   $(D_A D_B\alpha)^{\tf}$ vanishes.

The above analysis goes through if we replace $V^a$ by a local CKV so that $\xi^a$ represents a generator of the extended BMS group. In this case however, the soft term does not vanish. In $(z,\zb)$ coordinates the action takes the form:
\be
(X_{V})_{zz} = \L_V \sigma_{zz} - \alpha \sigma_{zz} +u \alpha \dot{\sigma}_{zz} - \frac{u}{2} D^3_z V^z ,\label{Xxiext} 
\ee
where we used the fact that $D_{z}D_{z} (D_{\zb}V^{\zb})=0$ for local CKV. Similar expression holds for the $\zb\zb$ component. In quantum theory, the action (\ref{Xxiext}) is generated by the charge $Q=Q_H+Q_S$ given in Eq. (5.10) of \cite{virasoro}.

\subsection{Mode functions}
In this section we describe the classical functions in radiative phase space that correspond to the standard creation/annihilation operators of gravitons in  quantum theory. These are essentially given by the $zz$ and $\zb \zb$ components of the Fourier transform of $\sigma_{AB}$, %We start by considering the Fourier transform in $u$ of $\sigma_{AB}$:
\be
\sigma_{AB}(\w,\xh) := \int_{-\infty}^\infty  \sigma_{AB}(u,\xh) e^{i \w u} du \label{sigmaw}.
\ee
As long as  $\w \neq 0$, (\ref{sigmaw})  admits a HVF \footnote{In a distributional sense;  strictly speaking one should integrate (\ref{sigmaw}) with a smearing function in $(\w,\xh)$ with support outside $\w =0$.} and hence we can find their PBs. The non vanishing  PBs are found to be:\footnote{In the present subsection as well as in section \ref{sec5},   $\gamma_{z \zb} \equiv q_{z \zb} = \sqrt{\gamma} = 2 (1 + z \zb)^{-2}$.} 
\be
\{ \sigma_{zz}(\w,z,\zb), \sigma_{\zb \zb}(\w',z',\zb') \} = \frac{\pi}{i \w} \sqrt{\gamma} \; \delta(\w+\w') \deltat(z-z') .\label{pbmode}
\ee
For later purposes, we note that the relation of the mode functions (\ref{sigmaw}) with the Fourier transform of the News tensor (\ref{news}) is given by:
\be
\sigma_{AB}(\w,\xh) = (2 i \w)^{-1} N_{AB}(\w,\xh). \label{sigmanewsw}
\ee

Following sections 5 of \cite{st} and 5.3 of \cite{virasoro}  (see also \cite{am,frolov,dp1}), we can find the relation of (\ref{sigmaw}) with the creation/annihilation functions from standard perturbative gravity. Following \cite{virasoro} we take coordinates in past null infinity that are antipodally related to those of future null infinity. In that case the expressions relating `in' quantities take the same form as the expressions relating  `out' quantities.  The following discussion thus applies to either case.   

The `annihilation function' $a_\pm(\w,\xh)$,  $\w >0$, of a  helicity $\pm 2$  graviton is found to be given by:   
\be
a_+(\w,\xh) = \frac{4 \pi i }{\sqrt{\gamma}} \sigma_{zz}(\w,\xh), \; \quad a_-(\w,\xh) = \frac{4 \pi i }{\sqrt{\gamma}} \sigma_{\zb\zb}(\w,\xh). \label{asigma1} 
\ee
Since $\overline{\sigma_{zz}}(\w)=\overline{\sigma}_{\zb \zb}(\w)=\sigma_{\zb \zb}(-\w)$, the relations for the `creation functions' have the opposite relation between helicity and holomorphic components:
\be
a_+(\w,\xh)^\dagger = -\frac{4 \pi i }{\sqrt{\gamma}} \sigma_{\zb \zb}(-\w,\xh), \; \quad a_-(\w,\xh)^\dagger = -\frac{4 \pi i }{\sqrt{\gamma}} \sigma_{z z}(-\w,\xh) ,\label{asigma2} 
\ee
where in the present classical context, the dagger  just means complex conjugation. The Poisson bracket (\ref{pbmode}) implies 
\be
\{ a_h (\w,\xh), a_{h'}(\w',\xh')^\dagger \} = \frac{2 (2 \pi )^3}{i \w \sqrt{\gamma}} \delta_{h h'} \delta(\w-\w') \delta(\xh,\xh') , \label{PBaa}
\ee
and corresponds to the Poisson brackets the functions have from the perspective of perturbative gravity: $\{a^h_{\vec{p}}, (a^{h'}_{\vec{q}})^\dagger \} = -i 2 E_{\vec{p}} \, \delta_{h h'} (2 \pi)^3   \delta^{(3)}(\vec{p}-\vec{q})$,  with $\vec{p}= \w \xh$ and $\vec{q}= \w' \xh'$.

\subsection{Action of BMS on mode functions} \label{sec4D}
The action of BMS on the mode functions can be obtained by taking the Fourier transform of (\ref{boost1}). Here we are interested in rotations and boost, so we  focus attention in the action of a BMS vector field of the form (\ref{boostrot}). Taking the Fourier transform of (\ref{XxiAB}) one finds:
\be
X_V (\sigma_{AB}(\w,\xh))= \L_{V} \sigma_{AB}(\w,\xh) - 2 \alpha \sigma_{AB}(\w,\xh) - \alpha \w \partial_\w \sigma_{AB}(\w,\xh). \label{boost2}
\ee
From (\ref{asigma1}), (\ref{asigma2}) one can verify that the corresponding action on the creation/annihilation functions is given by the differential operator:
\be
 J^h_V := V^z \partial_z + V^{\zb} \partial_{\zb}- \frac{1}{2} (D_z V^z + D_{\zb} V^{\zb})\w \partial_\w + \frac{h}{2} ( \partial_z V^z - \partial_{\zb} V^{\zb}) , \label{JVh}
\ee
according to
\be
X_V (a_h(\w,z,\zb) )  =  J^h_V a_h(\w,z,\zb) \ ; \ \quad X_V (a_h(\w,z,\zb)^\dagger )  =  J^{-h}_V a_h(\w,z,\zb)^\dagger . \label{Xah}
\ee
In quantum theory, $J^h_V$ represents the total angular momentum of a  helicity $h=\pm 2$ graviton. %In appendix \ref{spinorhelapp} we verify that $J^h_V$  coincides with the expression of total angular momentum  given in spinor-helicity variables. 
\section{Generalized BMS and radiative phase space} \label{sec4}
\subsection{Intrinsic characterization of generalized BMS group}
From the perspective of null infinity, the proposed generalized BMS vector fields $\xi^a$ are given by supertranslations and vector fields of the form (\ref{boostrot}) with the CKV condition on $V^A$ dropped.  The dropping of the CKV condition implies that $\xi^a$ does not preserve the universal structure $[(q_{ab},n^a)]$ described in section \ref{sec3.1}. It is natural to ask whether there is any other geometrical structure that is kept invariant under the action of generalized BMS. As we now show, such geometrical structure is given by equivalence classes of pairs $[(\e_{abc},n^a)]$ with $n^a$ as before, $\e_{abc}$ the volume form satisfying $\L_n \e_{abc}=0$, and equivalence relation given by
\be
(\e_{abc},n^a) \sim  (\w^3 \e_{abc},\w^{-1} n^a) ,\quad \forall \w: \I \to \reals : \L_n \w =0. \label{equiveps}
\ee
First, we notice that any generalized BMS vector field still satisfies $\L_{\xi}n^a= - \alpha n^a$, whereas its action on the volume form is \cite{AS,aabook}:
\begin{equation}
\begin{array}{lll}
{\cal L}_{\xi}\epsilon_{abc}\ =\ 3\alpha\epsilon_{abc},
\end{array}
\end{equation}
hence it keeps the pair $(\e_{abc},n^a)$ in the same equivalence class (\ref{equiveps}).  Conversely, one can verify that the group of symmetries of $[(\e_{abc},n^a)]$ is given by generalized BMS group. This can be shown along  the same lines as the proof given for the BMS case \cite{aabook}.  One finds that supertranslations are again a normal subgroup, and the quotient group is now the group of diffeomorphisms on the sphere.

\subsection{An example: action on radiative phase space of a massless scalar field} \label{sec5B}
One example of a radiative phase space where the underlying kinematical structure is  provided by the (equivalence class) of pairs $[(\epsilon_{abc}, n^{a})]$ is that of a massless scalar field \cite{AS}. As we show below, in this case it is indeed true that the generalized BMS group has a symplectic action.

The symplectic structure of the radiative phase space $\Gamma_\phi$ of a massless scalar field $\phi$ is given by \cite{AS}:
\be
\O_\phi(\phi_1,\phi_2)=\int d^3V (\phi_1 \L_n \phi_2 - \phi_2 \L_n \phi). \label{sympstrphi}
\ee

The symplectic structure (\ref{sympstrphi}) is defined in terms of the pair $(\e_{abc}, n^a)$ and there is a canonical isomorphism between different choices of pairs in the class (\ref{equiveps}) given by \cite{AS}: 
\be
(\e_{abc}, n^a) \to (\w^3 \e_{abc},\w^{-1} n^a), \quad  \phi \to \w^{-1} \phi . \label{isomphi}
\ee
The action of a generalized BMS vector field $\xi^a$ on $\Gamma_\phi$ can be obtained as in the BMS case for gravity discussed in section \ref{sec4B} and appendix \ref{boostapp}: First compute the variation of $\phi$ under  $\xi^a$ and then use the canonical isomorphism (\ref{isomphi}) to express the `transformed field' in the original `frame'. The result is:
\be
X^\phi_\xi = \L_\xi \phi + \alpha \phi \label{xiphi}.
\ee
The form (\ref{xiphi}) is the same as the one given in \cite{AS} for the action of BMS.  It is easy to verify that (\ref{xiphi}) is symplectic and that $[X_{\xi}, X_{\xi'}]=X_{[\xi,\xi']}$.

 \subsection{The case of gravitational radiative phase space} \label{sec4C}
Since generalized BMS does not preserve the universal structure $[(q_{ab},n^a)]$, and there is no (known to us)  natural isomorphism between the various universal structures  that generalized BMS can map to (namely those compatible with $[(\e_{abc},n^a)]$), we lack a geometrical framework from which we can attempt to  derive an action of generalized BMS on the radiative phase space of gravity. Thus, the strategy followed in sections \ref{sec4B} and \ref{sec5B} is not available. This problem is the phase-space counterpart of the issue discussed in section \ref{sec2.3}:  As generalized BMS vector fields change the leading order metric at ${\cal I}$, it is not clear  how to deduce an action of $\G$ on the free data.\\
We shall limit ourselves to present an ad-hoc HVF $X_\xi$.  The interest in this proposal lies in the fact that the associated ``Ward identities'' will be shown to be in one to one correspondence with the Cachazo-Strominger (CS) soft theorem.\\
There are however two shortcomings of our proposal which we hope to address in the future investigations.\\
\noindent{(1)}  The HVFs do not represent an action of generalized BMS since in general  $[X_{\xi},X_{\xi'}] \neq X_{[\xi,\xi']}$.\footnote{The situation is thus analogous to the recently proposed symmetries for massless QED that follow from the subleading soft photon theorem \cite{lowsym}.}% There also, the commutator of symmetry generators do not close among themselves.} 
\\
\noindent{(2)} The HVFs do not respect the fall-off behaviour of the radiative data and hence strictly speaking are not well defined on the entire phase space. (This infrared divergence is also present when the underlying vector fields are local CKVs.)

Our definition for the HVF is exactly the same as in (\ref{Xxiext}), where $V^{A}$ is an arbitrary (smooth) vector field on the sphere and $\alpha = (D_A V^A)/2$. It is sum of a hard and soft terms:
\be
X_V = \Xhom_V + \Xinhom_V,
\ee
where
\be
(\Xhom_V)_{zz}   =  \L_V \sigma_{zz} - \alpha \sigma_{zz} +u \alpha \dot{\sigma}_{zz} \label{Xhom} \ ; \ \quad (\Xinhom_V)_{zz}   :=   -\frac{u}{2} D^3_{z} V^z ,
\ee
and corresponding $z \to \zb$ expressions for  $(X_V )_{\zb \zb}$.
It can be seen that  $\Xhom_V$  preserves the fall-offs (\ref{falloffs}). Further, as shown in appendix \ref{sympapp}, it is symplectic: 
\be
\O(\Xhom_{V}(\sigma_1),\sigma_2) + \O(\sigma_1,\Xhom_{V}(\sigma_2)) =0 \quad  \forall \; \sigma_1, \sigma_2 \in  \Gamma. \label{svf}
\ee
Being linear in $\sigma$, its Hamiltonian can be found by:
\be
\Hhom_V(\sigma) := \frac{1}{2}\O(\Xhom_V(\sigma),\sigma),
\ee
which leads to  the same expression as the Hamiltonian for boosts (with the CKV condition on $V^A$ dropped).

Unless $D^3_{z} V^z =D^3_{\zb} V^{\zb}=0$, $\Xinhom_V$  diverges linearly in $u$ and hence  is not well defined on $\Gamma$. At a formal level $\Xinhom_V$ is however symplectic since it is just a c-number vector field.  We can make sense of the `would be' Hamiltonian on the subspace $\Gamma_0 \subset \Gamma$ given by:
\be
\Gamma_0 := \{ \sigma_{AB} \in \Gamma :  \sigma_{AB}(u,\xh) = \sigma^+(\xh) + O(u^{-1-\e}) \quad \text{as } \; u \to \pm \infty \} .
\ee
For $\sigma \in \Gamma_0$ we  define:
\be
\Hinhom_V(\sigma) :=  \Omega( \Xinhom_V,\sigma) =  - \int d^3 V  u \; (\dot{\sigma}^{zz} D^3_z V^z+ \dot{\sigma}^{\zb\zb} D^3_{\zb} V^{\zb}). \label{Hinhom}
\ee
\begin{comment}
It will be useful for later purposes to express  $\Hinhom(\sigma)$ in terms of the Fourier transform of the News tensor (\ref{news}),
\be
N_{AB}(\w,\xh)=  \int_{-\infty}^\infty  N_{AB}(u,\xh) e^{i \w u} du. \label{newsw}
\ee
We then have that $N_{AB}(\w=0,\xh) = -2 [\sigma_{AB}(\xh)]$ and
\be
\Hinhom_V(\sigma)  = i \partial_{\w} \oint d^2 V( N^{zz} D_z^3 V^z +N^{\zb \zb}D_{\zb}^3 V^{\zb} ) |_{\w=0}. \label{Hinhom2}
\ee
\end{comment}
Finally, for $\sigma \in \Gamma_0$  the total Hamiltonian is defined by
\be
H_V(\sigma) := \Hhom_V(\sigma) + \Hinhom_V(\sigma).
\ee
We will use these expressions to define the hard and soft operator in quantum theory.
In \cite{virasoro} $X_{V}$ is derived directly from the action of $V^{A}$ on $C_{AB}$ as given in Eq. (\ref{deltavcab}). If we follow this prescription here, it will lead to an expression for $X_{V}$ different from the one given above. However as the action of $\xi^a \partial_a = V^A \partial_A + u \alpha \partial_u$ changes the leading order metric at ${\cal I}$, this procedure is not applicable in this case.
%We conclude by emphasizing that the seemingly natural prescription for $X_{V}$ as given by eq.(\ref{deltavcab}) does not lead to the desired equivalence with the soft theorem.
\subsection{Action of generalized BMS generators on mode functions}
For $\w \neq 0$, the action of $X_V$ on the mode functions $\sigma_{AB}(\w,\xh)$ is fully determined by the term $\Xhom_V$. By taking the Fourier transform of (\ref{Xhom}) we arrive at the analogue of equation (\ref{boost2}) (with an additional `trace-free' symbol on the Lie derivative term). The corresponding action on the functions $a_{\pm}(\w,\xh)$ is given by the same equations as in the boost/rotation case, (\ref{Xah}), with the CKV condition on $V^A$ dropped. We thus find:
\be
\{ a_h(\w,z,\zb), H_V \}  =  J^h_V \, a_h(\w,z,\zb), \quad \{ a_h(\w,z,\zb)^\dagger, H_V \}  =  J^{-h}_V  \, a_h(\w,z,\zb)^\dagger ,\label{PBaV}
\ee
with $J^h_V$ the same differential operator given in Eq. (\ref{JVh}):  
\be
 J^h_V = V^z \partial_z + V^{\zb} \partial_{\zb}- \frac{1}{2} (D_z V^z + D_{\zb} V^{\zb})\w \partial_\w + \frac{h}{2} ( \partial_z V^z - \partial_{\zb} V^{\zb}) . \label{JVh2}
\ee
The non-closure of the HVFs $X_V$ manifests in a  particular simple form through the non-closure of the commutator of operators $J^h_V$  for general smooth vector fields.  A simple calculation reveals:
\be
[J^h_{V}, J^h_{W}] a_h = J^h_{[V,W]} \, a_h + h \, (\partial_{\zb} V^{z} \, \partial_z W^{\zb} - \partial_z V^{\zb}\, \partial_{\zb}W^{z}) \, a_h.
\ee
Thus,  the `non-closure' is proportional to the helicity. This is in accordance with the discussion of section \ref{sec5B}: The action of generalized BMS on the mode functions of a massless scalar field lacks a helicity contribution and  the non-closure term is absent there. 

\section{Generalized BMS and subleading soft theorem} \label{sec5}
In this section we show the equivalence between CS soft theorem and generalized BMS symmetries.  After summarizing the content of the soft theorem in section \ref{sec4A}, in section \ref{sec4A1} we propose the Ward identities for smooth vector fields belonging to the generalized BMS algebra. Although our derivation is simply a repeat of the derivation given in \cite{virasoro}, we express it in a slightly different form which facilitates the proof of the equivalence.\\
We then argue, in section \ref{sec4B1}, that the derivation of Ward identities associated to CS soft theorem as given in \cite{virasoro} goes through for smooth vector fields on the sphere. In section \ref{sec5.4} we show that using Ward identities for generalized BMS algebra, we can obtain the CS soft theorem. This derivation parallels the derivation for the case of supertranslations as mentioned in \cite{st}. We conclude in section \ref{sec5.5} with a comparison of this equivalence with the equivalence between Ward identities for supertranslations and Weinberg's soft graviton theorem.\\
%In section \ref{sec4A} we write the CS soft factor in terms of specific  generalized BMS vector fields. @@@as well as point out at magnetic eq zero? @@@ 

%In section \ref{sec4A} we display the form that a generalized BMS Ward identity would take in quantum theory. In section \ref{sec4B} we show that the CS subleading graviton theorem actually  implies these identities. In section \ref{sec4C} we show that conversely one can recover CS theorem from specific choices of generalized BMS vector fields.
In the following we work with the Fock space $\Fout$ generated by the standard creation/annihilation operators with nontrivial commutators given by $i$ times the  PBs (\ref{PBaa}):
\be
[ a^{\rm out}_h (\w,\xh), a^{\rm out}_{h'} (\w',\xh')^\dagger ] = \frac{2 (2 \pi )^3}{ \w \sqrt{\gamma}} \delta_{h h'} \delta(\w-\w') \delta(\xh,\xh') ,
\ee
and with the analogue $\Fin$ Fock space. The nature of the present section is rather formal. In particular, we do not construct the operator associated to  $H_V$ but rather assume that (i) it is normal ordered so that its action on the  vacuum is determined by the soft term;  (ii) its commutator with creation/annihilation operators is given by $i$ times the PBs (\ref{PBaV}).  Below we consider  `in' and `out' states of the form:
\be
\bra {\rm out} | := \bra 0 | \prod_{i=1}^{n_{\rm out}} a^{\rm out}_{h_i}(\Eout_i,\kho_i) \; , \;  \quad
| {\rm in} \ket := \prod_{i=1}^{n_{\rm in}} a^{\rm in}_{h_i}(\Ein_i,\khi_i)^\dagger | 0 \ket \label{inout}.
\ee

%It is important to  recall that the subleading ``soft" operator $\Hinhom$ that we have derived in the preceding section matches with $\hat{Q}_{\textrm{S}}$ as given in \cite{virasoro} only under 
The subleading soft operator which acts on asymptotic Fock states can be read off from Eq. (\ref{Hinhom}) and it precisely matches with the operator $Q_{S}^{\rm out}$ as given in \cite{virasoro}:
\begin{equation}
\begin{array}{lll}
(\Hinhom_V)^{\rm out}\ =\ \frac{1}{2}\int_{{\cal I}^{+}}dud^{2}z{D}_{z}^{3}V^{z}N^{z}_{\zb}\ =\ Q_{\textrm{S}}^{\textrm{out}}\ .
\end{array}
\end{equation}

\subsection{CS soft theorem} \label{sec4A}
In this section we summarize the content of CS soft theorem. We express the soft-factor in terms of a vector field on the sphere appearing in Eq. (6.6) of \cite{virasoro}. This will facilitate the discussions in the subsequent sections.
CS subleading soft theorem  for an outgoing soft graviton of helicity $h_s$ and momentum $q^\mu$
parametrized by $(\w,z_s,\zb_s)$ can be written as \cite{virasoro}:\footnote{The subsequent analysis can be easily extended to the case of incoming soft gravitons.}
\be
\lim_{\w \to 0^+} (1+ \w \partial_\w)   \bra {\rm out} | a^{\rm out}_{h_s} (\w,z_s,\zb_s) S | {\rm in} \ket  = \sum_{i} S^{(1) \, h_s}_i \bra {\rm out} |  S | {\rm in} \ket , \label{CSthm}
\ee
where the sum runs over all ingoing and outgoing particles. For an outgoing particle of momentum $k^\mu$ and helicity $h$ the soft factor is given by \cite{cs}:
\be
S^{(1) \, h_s}_{(k,h)} = (q \cdot k)^{-1} \e^{h_s}_{\mu \nu}(q) k^\mu q_\rho J^{\rho \nu}, \label{S1}
\ee
where $\e^{h_s}_{\mu \nu}(q)= \e^{h_s}_\mu(q) \e^{h_s}_\nu(q)$ is the polarization tensor of the soft graviton and $J^{\rho \nu}$ the total angular momentum of the $(k,h)$ particle. Following Strominger and collaborators, we seek to express (\ref{S1}) in holomorphic coordinates.  Let $(E,z,\zb)$ be the parametrization of the 4-momentum $k^\mu$. As discussed in section \ref{sec4D}, the total angular momentum can be expressed in terms of the differential operator $J^h_V$ given in Eq. (\ref{JVh}). The six CKVs corresponding to the  $(\mu,\nu)$ components are: 
\be
V^A_{i 0} :=  D^A \kh_i \, , \quad V^A_{i j} := \kh_i D^A \kh_j  - \kh_j D^A \kh_i , \quad i,j=1,2,3, \; A=z,\zb,
\ee
so that,
\be
J_{\mu \nu}  =  J^{h}_{V_{\mu \nu}}, \quad \mu,\nu=0,1,2,3 ,
\ee
where $\kh$ is the unit direction on the sphere parametrized by $(z,\zb)$. For the polarization tensor we follow \cite{st,virasoro} and take:
\be
\e^+(q)^\mu = \frac{1}{\sqrt{2}}(\zb_s,1,-i,-\zb_s) , \quad \e^-(q)^\mu = \frac{1}{\sqrt{2}}(z_s,1,i,-z_s) .\label{polvec}
\ee
Notice that (\ref{S1}) takes the form of a function of $(z,\zb)$ times a linear combination of boosts and rotations (with coefficients depending on  $z_s,\zb_s$ and $h_s$). Thus, all  $(z,\zb)$-independent factors multiplying $J^{\rho \nu}$ can be realized as linear combinations of CKVs. For instance:
\be
\e^{+}_{\nu}(q) q_\rho J^{\rho \nu} = J^{h}_{\e^{+}_{\nu}(q) q_\rho V^{\rho \nu}}.
\ee
Taking this into account, (\ref{S1}) can be written as:
\be
S^{(1) \, +}_{(k,h)} =  (z-z_s)^{-1} J^{h}_{(\zb-\zb_s)^2\partial_{\zb} } \, , \quad S^{(1) \, -}_{(k,h)} = (\zb-\zb_s)^{-1} J^{h}_{(z-z_s)^2 \partial_{z}}. \label{S12}
\ee
We finally show that (\ref{S12}) can in fact be written in terms of the  vector fields,
\be
K_{(+, \, z_s, \zb_s)}  :=  (z-z_s)^{-1}(\zb-\zb_s)^2\partial_{\zb} \, , \quad K_{(-, \, z_s, \zb_s)}   := (\zb-\zb_s)^{-1} (z-z_s)^2 \partial_{z} , \label{defK}
\ee
according to:
\be
S^{(1) \, +}_{(k,h)} =  J^{h}_{K_{(+, \, z_s, \zb_s)}} \, , \quad S^{(1) \, -}_{(k,h)} =  J^{h}_{K_{(-, \, z_s, \zb_s)}}. \label{S13}
\ee
Let us discuss the `$-$', case, the `$+$' one being analogous.   From the definition of $J^h_V$, (\ref{JVh2}) one can verify:
\ba
 J^{h}_{(\zb-\zb_s)^{-1} (z-z_s)^2 \partial_{z}}=  (\zb-\zb_s)^{-1} J^{h}_{(z-z_s)^2 \partial_{z}} + \frac{1}{2}( -E \partial_ E + h) (z_s-z)^2 \partial_z \frac{1}{(\zb-\zb_s)}.
\ea
The second term is proportional to $(z_s-z)^2 \delta^{(2)}(z,z_s)$. As long as (\ref{CSthm}) is understood as a distribution to be smeared against a smooth function on the sphere, this term vanishes and we obtain (\ref{S13}).

\subsection{Proposed Ward identities} \label{sec4A1}
In this section we motivate a proposal for the ``Ward identities".\footnote{The quotation marks are placed to remind us that the proposed charges do not yield a representation of the generalized BMS algebra on the radiative phase space.} This proposal is 
a straightforward generalization of the Ward identities proposed for the local CKVs  associated to the extended BMS algebra. We repeat the derivation here in the interest of pedagogy and for introducing notation for later  use.\\
Consider the analogue of the Virasoro symmetry proposed in  \cite{virasoro}, but with with $V^A$ a smooth vector field on the sphere  rather than a local CKV:
\be
\Hout_V S= S \Hin_V \label{HxiS} .
\ee
The evaluation of (\ref{HxiS}) between the states (\ref{inout}) is obtained by using the commutators (see Eq. (\ref{PBaV})):
\be
[a^{\rm out}_h(\w,\xh),\Hout_V]  =  i  J^{\; h}_V a^{\rm out}_h(\w,\xh) , \quad [ a^{\rm in}_h(\w,\xh)^\dagger,\Hin_V ] = i J^{-h}_V a^{\rm in}_h(\w,\xh) ,  \label{commaR}
\ee
together with the action  $H^{\rm in (out)}_V$  on the in (out) vacuum. This action is determined by the soft part of $H^{\rm in (out)}_V$ (\ref{Hinhom}). Following \cite{virasoro}, we express (\ref{Hinhom}) in terms the Fourier transform of the News tensor so that (the prescription for the $\w \to 0$ limit is described below):
\begin{multline}
\Hin_V | 0 \ket =\\
\hspace*{0.1in}-\frac{i}{2}\lim_{\w \to 0}  \partial_{\w} \oint d^2 V( N^{\rm in}_{z z}(\w,\xh) D^{z} D^{z} D_{\zb} V^{\zb}) + N^{\rm in}_{\zb\zb}(\w,\xh) D^{\zb} D^{\zb} D_z V^z| 0 \ket , \\ \label{Hinvac}
\end{multline}
and similar expression for $\bra 0 | \Hout_V$.
The matrix element of  (\ref{HxiS}) between the `in' and `out' states implies then:
\begin{multline}
 \frac{1}{2}\lim_{\w \to 0} \partial_\w \oint d^2 V D^{z} D^{z} D_{\zb} V^{\zb}\\
 \hspace*{0.5in}\left( \bra {\rm out} | N^{\rm out}_{z z}(\w,\xh) S | {\rm in} \ket - \bra {\rm out} | S N^{\rm in}_{z z}(\w,\xh) | {\rm in} \ket \right) + z \leftrightarrow \zb = \\  \sum_i J^{h_i}_{V_i} \bra {\rm out} |  S | {\rm in} \ket. \label{HxiSme}
\end{multline}
The sum runs over all `in' and `out' particles, with  the convention that for an  `in' particle one takes  $J^{h_i}_{V_i} = J^{-h^{\rm in}_i}_{V_i}$  according to (\ref{commaR}).     

We now focus on the LHS of (\ref{HxiSme}).   Firstly we need to specify how the $\w \to 0$ limit is taken. We take $\w \to 0^+$ in (\ref{HxiSme}) so that only the `out' term survives. This prescription is slightly different than the one  given in \cite{virasoro}. However it leads to the same form of Ward identities as given in \cite{virasoro}.\footnote{For superstranslations, this prescription also leads to the same Ward identities of \cite{st}.} 

With this prescription, and using Eqns. (\ref{sigmanewsw}), (\ref{asigma1}) the LHS of (\ref{HxiSme})  takes the form:
\be
\lhs =  \frac{1}{4 \pi} \lim_{\w \to 0^+} (1+ \w \partial_\w) \int d^2 z ( D^3_{\zb}  V^{\zb}  \bra {\rm out} | a^{\rm out}_+(\w,\xh) S   \ket + D^3_{z} V^z \bra   a^{\rm out}_-(\w,\xh) S | {\rm in}  \ket ) \label{lhsward}
\ee
where we used  $\sqrt{\gamma}\sqrt{\gamma}\gamma^{z \zb}\gamma^{z \zb}=1$. Substituting Eq. (\ref{lhsward}) in Eq. (\ref{HxiSme}) we obtain the proposed  identities. They take precisely the same form as the Virasoro Ward identities of  \cite{virasoro} : 
\begin{multline}
 \frac{1}{4 \pi} \lim_{\w \to 0^+} (1+ \w \partial_\w)\\
 \hspace*{0.4in}\int d^2 z ( D^3_{\zb}  V^{\zb}  \bra {\rm out}| a^{\rm out}_+(\w,\xh) S  | {\rm in} \ket + D^3_{z} V^z \bra {\rm out}|  a^{\rm out}_-(\w,\xh) S | {\rm in} \ket ) = \\  \sum_i J^{h_i}_{V_i} \bra {\rm out} |  S | {\rm in} \ket.  \label{wardid}
\end{multline}

\subsection{From CS theorem to  generalized BMS symmetries}\label{sec4B1}
The purpose of this section is to show that remarkably enough, the derivation of the Virasoro Ward identities given in \cite{virasoro} does not make use of the CKV property of the  vector fields in question,  so that the identities  hold for arbitrary smooth vector field on the sphere.\footnote{In fact, due to their singular nature, it is  not clear to us how the derivation works for local CKVs.} 

%We will now show that when we interpret these Ward identities as associated to generalized BMS (rather then the extended one) it leads to the proposed equivalence with CS soft theorem.\\ 

From Eqns. (\ref{CSthm}), (\ref{S13}), CS theorem can be written as: 
\be
\lim_{\w \to 0^+} (1+ \w \partial_\w)   \bra {\rm out} | a^{\rm out}_{h_s} (\w,z,\zb) S | {\rm in} \ket  = \sum_{i} J^{h_i}_{K^i_{(h_s, \, z, \zb)}} \bra {\rm out} |  S | {\rm in} \ket . \label{CSthm2}
\ee
 Let $V^A \partial_A$ be any smooth vector field on the sphere. In the following we work with $V^z \partial_z$ and $V^{\zb} \partial_{\zb}$ components separately. Multiplying  the LHS of  Eq. (\ref{CSthm2}) with $h_s=-2$ by  $(4\pi)^{-1}  D^3_z V^z $ and integrating over $(z,\zb)$, we obtain the LHS of the  proposed Ward identity (\ref{wardid}) for the vector $V^z \partial_z$. The same operation on the RHS of (\ref{CSthm2}) is given by:
\be
 (4 \pi)^{-1} \sum_{i} \int d^2 z  D^3_z V^z J^{h_i}_{K^i_{(-, \, z, \zb)}} \bra {\rm out} |  S | {\rm in} \ket  = \sum_i J^{h_i}_{W_i}\bra {\rm out} |  S | {\rm in} \ket \label{SW}
\ee
where
\be
W_i := (4 \pi)^{-1} \int d^2 z  D^3_z V^z {K^i_{(-, \, z, \zb)}} . \label{Wi}
\ee
In order to integrate by parts in (\ref{Wi}), we need to specify the tensor index structure of  $K^i_{(-, \, z, \zb)}$  with respect to the $(z,\zb)$ coordinates. This tensor structure is given by  $a^{\rm out}_- (\w,z,\zb) \sim \sigma_{\zb\zb}/\sqrt{\gamma}$ due to Eqns. (\ref{asigma1}), (\ref{CSthm2}). Following \cite{virasoro}, this is captured by $\eh_{\zb \zb} := \sqrt{\gamma}$. We thus obtain  (to avoid confusion we set $K^i_- \equiv K^i_{(-, \, z, \zb)}$):
\begin{multline}
\int d^2 z D^3_z V^z K^{i}_{-}  =  \int d^2 z \sqrt{\gamma} \gamma^{z \zb}\gamma^{z \zb} D_z D_z (D_z V^z) (\eh_{\zb \zb} K^{i}_{-}) = \\
 -  \int d^2 z \sqrt{\gamma} V^z D_z D^{\zb} D^{\zb} (\eh_{\zb \zb} K^{i}_{-}) = 4 \pi V^{z_i}(z_i,\zb_i) \partial_{z_i}  ,
\end{multline}
where in the last equation we used an identity given in Eq. (6.7) of \cite{virasoro}: 
\be
\gamma^{z \zb} D^3_z( \eh_{\zb \zb} K^{i}_{-}) = -4 \pi \delta^{(2)}(z-z_i) \partial_{z_i}.
\ee
Using this result back in (\ref{SW}) we recover the RHS of the proposed Ward identity (\ref{wardid}) for the vector $V^z \partial_z$. Similar discussion applies for $h_s=+2$ and the vector $V^{\zb} \partial_{\zb}$.  Adding the two results one obtains the Ward identity (\ref{wardid}) for the vector field $V^A \partial_A$.

\subsection{From Ward identity to soft theorem} \label{sec5.4}
CS theorem can be recovered as the Ward identity associated to the vector fields (\ref{defK}).\footnote{As in the case of supertranslations, this derivation requires a choice of a non-smooth (in the present case $C^{1}$) vector field. It is understood that this is due the use of sharp momentum eigenstates.}  For the  case of an outgoing negative helicity soft graviton with direction $(z_s, \zb_s)$, we choose $V^A$ in (\ref{wardid}) by
\be
V^A  \partial_A = K_{(-, \, z_s, \zb_s)}  =  (\zb-\zb_s)^{-1}(z-z_s)^2\partial_{z}.
\ee
One can verify that 
\be
D^3_{z} K_{(-, \, z_s, \zb_s)}^{z} = 4 \pi \delta^{2}(z-z_s). \label{D3K}
\ee
 Using (\ref{D3K}) in (\ref{wardid}) we recover CS theorem (\ref{CSthm2}) for $h_s=-2$. Similar discussion applies for a positive helicity soft graviton.\\
\subsection{Comparison with supertranslation case} \label{sec5.5}
We now note the following subtlety regarding this equivalence.  Recall that   Weinberg's soft graviton theorem is equivalent to the Ward identities associated  to the supertranslation symmetries \cite{st}. As supertranslations are parametrized by a single function, it is rather surprising that the associated Ward identities can give rise to the soft graviton theorem for both positive as well as negative helicity soft particles. That this is possible is due to a so called global constraint which underlie the definition of CK spaces. On future null infinity, it is given by:
\begin{equation}
[D_{z}^{2}C_{\zb\zb}\ -\ D_{\zb}^{2}C_{zz}]_{{\cal I}^{+}_{\pm}}\ =\ 0
\end{equation}
It can be re-written in terms of the zero mode of the News tensor as
\begin{equation}
D_{z}^{2}N_{\zb\zb}^{\textrm{out}}(\omega=0,\hat{x})\ =\ D_{\zb}^{2}N_{zz}^{\textrm{out}}(\omega=0,\hat{x}).
\end{equation}
This constraint ensures that the operator insertions due to positive and negative helicity soft gravitons are equivalent to each other.  (For more details we refer the reader to \cite{strominger2}.)  This is consistent with the remarkable structure of Weinberg's soft term which does not depend on the angular momenta of the external particles.

However this constraint does not imply that the operator insertions associated to ``subleading" soft positive helicity gravitons (i.e. when the leading order pole is projected out from the insertion) are equivalent to those of negative helicity gravitons. This is consistent with the fact that this subleading theorem is equivalent to Ward identity associated to vector fields on a sphere which are parametrized by two independent functions. This is in turn  reflected  in the structure of the sub-leading CS soft term which depends on the angular momenta of the scattering particles. 

%However this constraint does not imply that the operator insertions associated to ``subleading" soft positive helicity gravitons (i.e. when the leading order pole is projected out from the insertion) are equivalent to those of negative helicity gravitons. This is consistent with the fact that this subleading theorem is equivalent to Ward identity associated to vector fields on a sphere which are parametrized by two independent functions. 

\section{Outlook}
Motivated by the desire to understand the subleading soft graviton theorem as arising from Ward identities associated to  asymptotic symmetries, we considered a \emph{distinct} generalization of the BMS group than the one proposed in \cite{barnich1}. We showed that $\G$, which is essentially obtained by dropping a single condition from the definition of the BMS group (namely that the vector fields defined on the conformal sphere be CKVs) is a semi-direct product of supertranslations and diffeomorphisms of the conformal sphere, $\G = \textrm{ST} \rtimes \textrm{Diff}(S^{2})$.   We argued that $\G$ acts as a symmetry group on the space of all asymptotically flat geometries which are in a suitable neighborhood of  Minkowski space-time. \\

Associated to vector fields which generate $\textrm{Diff}(S^{2}) = \G/\textrm{{ST}}$  we  proposed a definition of the flux in the radiative phase space of Ashtekar which was motivated by the definition of corresponding flux for the Virasoro symmetries. The reason why we have not been able to derive this flux expression from first principles (as one can do for any vector field belonging to extended BMS) can be most easily understood as follows.\footnote{For pedagogy we restrict our attention to future null infinity.} \\

In the case of Virasoro symmetries, the derivation of flux in the radiative phase space is based upon the action of extended BMS vector fields on the free data quantified by $C_{AB}$ \cite{virasoro}. $C_{AB}$ is the free (radiative) data in the sense that  it is unconstrained and that all the other dynamical metric components  in the neighborhood of null infinity are determined from Einstein's equations using $C_{AB}$. However what constitutes the free data is ``frame dependent" in the sense that it depends on the chosen `kinematical',  leading order  metric at null infinity. As extended BMS group preserves the leading order metric at ${\cal I}^{+}$, it maps a given radiative data into a different radiative data. 
Due to the fact that $\G$ changes the leading order structure of the metric components we have been unable to derive the action of its proposed flux from first principles. However we think that its appeal lies in the fact that the related Ward identities are equivalent to the subleading soft graviton theorem.\\

Yet another unresolved issue with $H_{V}$ (as is also the case for  new class of asymptotic symmetries proposed for massless QED \cite{lowsym}) is that the fluxes associated to $\G$ do not form a closed algebra. It is conceivable that this is due to the fact that the radiative phase space of Ashtekar is based upon the existence of a fixed kinematical structure (namely the conformal metric on the sphere and the null vector field $n^{a}$) which is in turn tied to the existence of a fixed space-time metric at leading order in $r$. This expectation is borne out by the fact that in the case of massless scalar field where the radiative phase space does not refer to the entire conformal metric but only the volume form, these symmetries do indeed form a closed algebra.\footnote{Note that if this expectation turned out to be true, then both the issues mentioned above are two sides of the same coin.}\\

In light of what is said above, there  appear certain natural directions in which a systematic derivation of the fluxes associated to $\G$ (such that they form a closed algebra) could be obtained, namely by weakening the dependence of radiative phase space on the universal structure. Detailed implementation of this idea is currently under investigation.\\

In summary our proposal for $\G$ as a group of asymptotic symmetries for low energy gravitational scattering processes is at best a tentative one. However due to its relationship with the subleading soft theorem we believe that  further investigation of the above mentioned issues is warranted.\\

\noindent {\bf Acknowledgements}\\
We are indebted to Abhay Ashtekar for stimulating discussions and suggestions. We are grateful to A. P. Balachandran and Sachindeo Vaidya for insightful discussions associated to asymptotic symmetries in gauge theories. We would also like to thank the participants of the workshop ``Asymptotia" held at Chennai Mathematical Institute for many discussions related to BMS group. We thank Burkhard Schwab and an anonymous referee for their comments on the manuscript. AL is supported by Ramanujan Fellowship of the Department of Science and Technology. 
\appendix
\section{Zero mode subtleties of Poisson brackets} \label{PBapp}
Since the subtleties we want to discuss  arise from the dependance in $u$, in this appendix we suppress the angular components and take $\sigma$ to be  a scalar function on the real line parametrized by $u$. More precisely, we consider the phase space  $\Gamma$ of scalar functions on the real line with fall-offs $\sigma = \sigma^{\pm}+ O(u^{-\e})$ as $u \to \pm \infty$ and symplectic form:
\be
\O(\s_1,\s_2)= \int  ( \s_1 \dot{\s}_2 - \dot{\s}_1 \s_2 ) d u = 2 \int \s_1 \dot{\s}_2 du  -  [\s_1 \s_2] , \label{wapp}
\ee
where the square bracket denote difference in evaluation at $u = \pm \infty$. 

Consider a phase space function of the form:
\be
F(\sigma):= \int  F(u) \sigma(u) du ,\label{Fsigma}
\ee
for some smearing function $F(u)$. To find the corresponding HVF $f:=X_F$, we need to solve the equation  
\be
F(\sigma) = \O(f, \sigma) \label{Ff}
\ee
for  $f \in \Gamma$. From (\ref{wapp}) it follows that we should have $F=-2 \dot{f}$ and $[f \sigma]=0 \; \forall \sigma$. The condition involving the boundary term can only be satisfied if $f^+= f^-=0$. The two conditions can be summarized by:
\be
\begin{array}{ll}
i) &    F \sim 1/|u|^{1+\epsilon} \quad {\rm as} \; u \to \pm \infty \\
 ii) & \int_{-\infty}^{\infty} F du =0. 
  \end{array} \label{condF}
 \ee
Only for $F$ satisfying i) and ii) does  (\ref{Fsigma}) admits a HVF, in which case it is given by:
\be
X_F = f(u)= -\frac{1}{2} \int_{-\infty}^{u} F(u') du'.
\ee
The  PB between a pair of functions $F$ and $G$ satisfying (i) and (ii) can then be written as:
\ba
\{F,G \} & = & \O(X_F,X_G) \\
&=&- \frac{1}{2}\int_{-\infty}^{\infty} du du' G(u) F(u') \theta(u-u') \\
&=& -\frac{1}{4}\int_{-\infty}^{\infty} du du' G(u) F(u') \sign(u-u') \label{stpb}
\ea
where $\theta(u-u')$ is the step function. It is clear that there is no way to extract PBs for $\{\sigma(u),\sigma(u') \}$ (there is not even a unique expression).   Let us nevertheless use the form (\ref{stpb}) to set,
\be
 \text{``} \{\sigma(u),\sigma(u') \}= -\frac{1}{4}\sign(u-u') \; \text{''} \label{stpb2},
 \ee
and see how we get a contradiction. Eq. (\ref{stpb2}) is the analogue of Eq. (2.12) of \cite{st} (our conventions for PBs differ by a sign with those used in \cite{st}). An example of the contradiction found in \cite{st} is as follows. Consider the phase space function
\be
H(\sigma) :=[\sigma]=\sigma^+ - \sigma^- =  \O(1,\sigma).
\ee
It admits a HVF given by  $X_H = 1$. Its action on $\sigma(u)$ is simply given by $X_H(\sigma(u))=1$.  Since  $\sigma(u)$ does not admit a HVF (not even in a distributional sense), we cannot interpret this action in terms of Poisson brackets.  If we nevertheless do so, we  find:
\be
\text{``} \,  \{ \sigma(u), [\sigma] \}  \, \text{''}  = \;  X_H(\sigma(u)) = 1 .
\ee
But using (\ref{stpb2}) we get
\be
 \text{``} \{ \sigma(u), [\sigma] \} = \{ \sigma(u), \s^+ \} - \{ \sigma(u), \s^- \} = \frac{1}{4} - (-\frac{1}{4}) = \frac{1}{2}  \; \text{''},
\ee
and hence the contradiction.

\section{Derivation of action of BMS  on radiative phase space} \label{boostapp}
In \cite{AS} it is shown that given a derivative $D_a$ compatible with $(q_{ab},n^a)$ and a new frame $(q'_{ab},n'^a)=(\w^2 q_{ab},\w^{-1} n^a)$ there exists a natural derivative $D'_a$ compatible with the new frame given by  Eq. (4.5) of \cite{AS}:  
\be
D'_a k_b = D_a k_b - 2 \w^{-1} k_{(a} \partial_{b)} \w + \w^{-1}q_{ab} \w^c k_c ,\label{newD}
\ee
where $\w^c$ is any vector satisfying $\w^c q_{bc}= D_b \, \w$.  The corresponding map $[D_a] \to [D'_a]$ between equivalence classes of derivatives provides the isomorphism between the phase spaces associated to the two frames. 

Under the action of a general BMS vector field $\xi^a$, the `transformed derivative'  $D'_a \approx D_a +\delta_\xi D_a $ is compatible with   the frame $(q'_{ab}, n'^a) \approx ((1+ 2\alpha) q_{ab}, (1-\alpha) n^a)$. To obtain the BMS action on the original phase space, we use the aforementioned isomorphism to map $D'_a$ to a derivative compatible   $(q_{ab},n^a)$. The resulting derivative,   $D''_a$, is obtained  by performing the substitutions   $D'_a \to D''_a$,  $D_a \to  D'_a$, and $\w \to 1- \alpha$ in (\ref{newD}):
\be\label{A2}
D''_a k_b - D_a k_b = (\delta_\xi D_a) k_b + 2(1+\alpha)  k_{(a} \partial_{b)} \alpha - (1+\alpha) (2+\alpha) q_{ab} q^{cd}l_c \partial_d \alpha. 
\ee
Choosing $k_b$ such that $n^b k_b=1$ and taking the trace free part,  (\ref{A2}) gives us the  desired action:
\be\label{A3}
(X_\xi)_{ab} =  ([\L_\xi,D_a] k_{b} +2 k_{(a} \partial_{b)} \alpha)^{\tf}
\ee
where the contribution of the last term in Eq. (\ref{A2}) is zero as it is pure trace and we have dropped $O(\alpha^{2})$ terms. Eq. (\ref{A3}) precisely matches with the Hamiltonian vector field associated to a BMS vector field $\xi^a$ as given in Eq. (4.14) of \cite{AS}.

\section{Symplectic action of generalized BMS generators} \label{sympapp}
 The proof of (\ref{svf}) is essentially the same  to that for BMS generators. The difference with the BMS case is that $\L_{V} q_{AB}$ contains trace-free components, which we denote by $t_{AB}$:
\be
\L_V q_{AB} = 2 \alpha q_{AB} + t_{AB} , \quad \text{where} \quad q^{AB}t_{AB}=0. \label{liexiq}
\ee
Similarly:
\be
\L_V q^{AB} = -2 \alpha q^{AB} - t^{AB}, \label{liexiqinv}
\ee
with $t^{AB}= t_{CD} q^{AC} q^{BD}$. That these `non-CKV' terms do not spoil (\ref{svf}) will follow from the fact that they will always appear contracted with two other trace-free tensors and a metric, yielding a vanishing result. For instance, if $\s^1_{AB}, \s^2_{AB}$ and $t^{AB}$ are symmetric and trace-free, then
\be
q^{AC} t^{BD} \s^1_{AB} \s^2_{CD} =0 , \label{qtss}
\ee
as can be seen by writing the expression in $(z,\zb)$ components.
 
  The expression (\ref{Xhom}) for  $\Xhom_V$ is:
 \be
(\Xhom_V)_{AB} = (\L_V \sigma_{AB})^{\tf}+ \alpha u \dot{\sigma}_{AB} - \alpha \sigma_{AB} , \label{Xhom2} 
\ee
where $\tf$ denotes trace free part with respect to $q_{AB}$. The evaluation of (\ref{svf}) involves three terms associated to each of the terms in (\ref{Xhom2}):
\be
\O(\Xhom_{V}(\sigma_1),\sigma_2) =\O((\L_{V} \sigma_1)^{\tf},\sigma_2) +\O(\alpha u \dot{\s}_1,\s_2)- \O(\alpha \s_1,\s_2) .
\ee
The first contribution to (\ref{svf}) is:
\ba
\O((\L_{V} \sigma_1)^{\tf},\sigma_2) - 1 \leftrightarrow 2 & = &  \int du d^2 Vq^{AC} q^{BD}( \L_V \s^1_{AB} \dot{\s}^2_{BC}  -  \L_V\dot{\s}^1_{AB} \s^2_{CD} ) - 1 \leftrightarrow 2  \nonumber \\
& = & \int du d^2 V q^{AC} q^{BD}(\L_V( \s^1_{AB} \dot{\s}^2_{BC})  -  \L_V(\dot{\s}^1_{AB} \s^2_{CD})) \nonumber \\
& =  &2  \int du d^2 V \alpha q^{AC} q^{BD}( \s^1_{AB} \dot{\s}^2_{BC}  - \dot{\s}^1_{AB} \s^2_{CD}) . \label{firstterm}
\ea
where we used that $q^{AC} q^{BD} (\L_V \s^1_{AB})^\tf \dot{\s}^2_{BC} =q^{AC} q^{BD} \L_V \s^1_{AB} \dot{\s}^2_{BC}$,  $\L_V (d^2 V) = 2 \alpha d^2 V$, and Eqns. (\ref{liexiqinv}), (\ref{qtss}). % For the reasons mentioned in Eq. (\ref{qtss}) there is no contribution from the trace-free terms $t^{AB}$.
The second contribution to (\ref{svf}) is:
\ba
\O(\alpha u \dot{\sigma}_1,\sigma_2) - 1 \leftrightarrow 2 & = &  \int du d^2 Vq^{AC} q^{BD}( \alpha u \dot{\s}^1_{AB} \dot{\s}^2_{BC}  -  \alpha \partial_u( u \dot{\s}^1_{AB})\s^2_{CD} ) - 1 \leftrightarrow 2  \nonumber \\
& =&  \int du d^2 Vq^{AC} q^{BD}(    \alpha ( u \dot{\s}^1_{AB}) \dot{\s}^2_{CD} ) - 1 \leftrightarrow 2 =0,
\ea
where we used that $\lim_{u \to \pm \infty} u \dot{\sigma}_{AB}=0$ so that no boundary contribution arises from the integration by parts in $u$.
Finally, it is easy to see that last contribution to (\ref{svf}) exactly cancels (\ref{firstterm}).


\begin{thebibliography}{99}
\bibitem{bms} 
  H.~Bondi, M.~G.~J.~van der Burg and A.~W.~K.~Metzner,
  ``Gravitational waves in general relativity. 7. Waves from axisymmetric isolated systems,''
  Proc.\ Roy.\ Soc.\ Lond.\ A {\bf 269}, 21 (1962).

\bibitem{sachs} 
  R.~K.~Sachs,
  ``Gravitational waves in general relativity. 8. Waves in asymptotically flat space-times,''
  Proc.\ Roy.\ Soc.\ Lond.\ A {\bf 270}, 103 (1962).
  
 \bibitem{aaprl} 
  A.~Ashtekar,
  ``Asymptotic Quantization of the Gravitational Field,''
  Phys.\ Rev.\ Lett.\  {\bf 46}, 573 (1981)

\bibitem{AS} 
  A.~Ashtekar and M.~Streubel,
  ``Symplectic Geometry of Radiative Modes and Conserved Quantities at Null Infinity,''
  Proc.\ Roy.\ Soc.\ Lond.\ A {\bf 376}, 585 (1981)

\bibitem{aajmp} 
  A.~Ashtekar,
  ``Radiative Degrees of Freedom of the Gravitational Field in Exact General Relativity,''
  J.\ Math.\ Phys.\  {\bf 22}, 2885 (1981)

\bibitem{aaoxf}
A.~Ashtekar,
``Quantization of the Radiative Modes of the Gravitational Field'', In: Quantum
Gravity 2; Edited by C. J. Isham, R. Penrose, and D. W. Sciama (Oxford University Press,
Oxford, 1981).


\bibitem{aabook} 
  A.~Ashtekar,
  ``Asymptotic Quantization'',
  Naples, Italy: Bibliopolis (1987) 

  \bibitem{kf} 
  P.~P.~Kulish and L.~D.~Faddeev,
  ``Asymptotic conditions and infrared divergences in quantum electrodynamics,''
  Theor.\ Math.\ Phys.\  {\bf 4}, 745 (1970)
 
 \bibitem{akhoury} 
  J.~Ware, R.~Saotome and R.~Akhoury,
  ``Construction of an asymptotic S matrix for perturbative quantum gravity,''
  JHEP {\bf 1310}, 159 (2013)
  [arXiv:1308.6285 [hep-th]]

 
 \bibitem{strominger1} 
  A.~Strominger,
  ``Asymptotic Symmetries of Yang-Mills Theory,''
  JHEP {\bf 1407}, 151 (2014)
  [arXiv:1308.0589 [hep-th]]

\bibitem{strominger2} 
  A.~Strominger,
  ``On BMS Invariance of Gravitational Scattering,''
  JHEP {\bf 1407}, 152 (2014)
  [arXiv:1312.2229 [hep-th]]

\bibitem{st} 
  T.~He, V.~Lysov, P.~Mitra and A.~Strominger,
  ``BMS supertranslations and Weinberg's soft graviton theorem,''
  arXiv:1401.7026 [hep-th]

\bibitem{weinberg} 
  S.~Weinberg,
  ``Infrared photons and gravitons,''
  Phys.\ Rev.\  {\bf 140}, B516 (1965)

    
  \bibitem{barnich1} 
  G.~Barnich and C.~Troessaert,
  ``Symmetries of asymptotically flat 4 dimensional spacetimes at null infinity revisited,''
  Phys.\ Rev.\ Lett.\  {\bf 105}, 111103 (2010)
  [arXiv:0909.2617 [gr-qc]
  
  
  \bibitem{barnich2} 
  G.~Barnich and C.~Troessaert,
  ``Aspects of the BMS/CFT correspondence,''
  JHEP {\bf 1005}, 062 (2010)
  [arXiv:1001.1541 [hep-th]]
  
  \bibitem{barnich3} 
  G.~Barnich and C.~Troessaert,
  ``BMS charge algebra,''
  JHEP {\bf 1112}, 105 (2011)
  [arXiv:1106.0213 [hep-th]]

  
  \bibitem{virasoro} 
 D.~Kapec, V.~Lysov, S.~Pasterski and A.~Strominger,
  ``Semiclassical Virasoro symmetry of the quantum gravity $ \mathcal{S}$-matrix,''
  JHEP {\bf 1408}, 058 (2014)
  [arXiv:1406.3312 [hep-th]].

  
  \bibitem{cs} 
  F.~Cachazo and A.~Strominger,
  ``Evidence for a New Soft Graviton Theorem,''
  arXiv:1404.4091 [hep-th].

\bibitem{plefka} 
 J.~Broedel, M.~de Leeuw, J.~Plefka and M.~Rosso,
  ``Constraining subleading soft gluon and graviton theorems,''
  Phys.\ Rev.\ D {\bf 90}, 065024 (2014)
  [arXiv:1406.6574 [hep-th]]
\bibitem{bern} 
 Z.~Bern, S.~Davies, P.~Di Vecchia and J.~Nohle,
  ``Low-Energy Behavior of Gluons and Gravitons from Gauge Invariance,''
  Phys.\ Rev.\ D {\bf 90}, no. 8, 084035 (2014)
  [arXiv:1406.6987 [hep-th]]

\bibitem{gj} 
  D.~J.~Gross and R.~Jackiw,
  ``Low-Energy Theorem for Graviton Scattering,''
  Phys.\ Rev.\  {\bf 166}, 1287 (1968)

\bibitem{naculich} 
  S.~G.~Naculich and H.~J.~Schnitzer,
  ``Eikonal methods applied to gravitational scattering amplitudes,''
  JHEP {\bf 1105}, 087 (2011)
  [arXiv:1101.1524 [hep-th]].


\bibitem{white} 
  C.~D.~White,
  ``Factorization Properties of Soft Graviton Amplitudes,''
  JHEP {\bf 1105}, 060 (2011)
  [arXiv:1103.2981 [hep-th]]
  
  \bibitem{sterman} 
  R.~Akhoury, R.~Saotome and G.~Sterman,
  ``Collinear and Soft Divergences in Perturbative Quantum Gravity,''
  Phys.\ Rev.\ D {\bf 84}, 104040 (2011)
  [arXiv:1109.0270 [hep-th]].
  
  \bibitem{beneke} 
  M.~Beneke and G.~Kirilin,
  ``Soft-collinear gravity,''
  JHEP {\bf 1209}, 066 (2012)
  [arXiv:1207.4926 [hep-ph]].
  
  
  \bibitem{bal1} 
  A.~P.~Balachandran and S.~Vaidya,
  ``Spontaneous Lorentz Violation in Gauge Theories,''
  Eur.\ Phys.\ J.\ Plus {\bf 128}, 118 (2013)
  [arXiv:1302.3406 [hep-th]]
  
  \bibitem{bal2} 
  A.~P.~Balachandran, S.~Kurkcuoglu, A.~R.~de Queiroz and S.~Vaidya,
  ``Spontaneous Lorentz Violation: The Case of Infrared QED,''
  arXiv:1406.5845 [hep-th]
  
  \bibitem{schwab} 
B.~U.~W.~Schwab and A.~Volovich,
  ``Subleading soft theorem in arbitrary dimension from scattering equations,''
  Phys.\ Rev.\ Lett.\  {\bf 113}, 101601 (2014)
  [arXiv:1404.7749 [hep-th]]

\bibitem{afkami} 
  N.~Afkhami-Jeddi,
  ``Soft Graviton Theorem in Arbitrary Dimensions,''
  arXiv:1405.3533 [hep-th]

\bibitem{zlotnikov} 
  M.~Zlotnikov,
  ``Sub-sub-leading soft-graviton theorem in arbitrary dimension,''
  JHEP {\bf 1410}, 148 (2014)
  [arXiv:1407.5936 [hep-th]]
  
  \bibitem{rojas} 
  C.~Kalousios and F.~Rojas,
  ``Next to subleading soft-graviton theorem in arbitrary dimensions,''
  arXiv:1407.5982 [hep-th]
  
   \bibitem{skinner} 
  T.~Adamo, E.~Casali and D.~Skinner,
  ``Perturbative gravity at null infinity,''
  Class.\ Quant.\ Grav.\  {\bf 31}, 225008 (2014)
  [arXiv:1405.5122 [hep-th]].
  
  \bibitem{mason} 
  Y.~Geyer, A.~E.~Lipstein and L.~Mason,
  ``Ambitwistor strings at null infinity and subleading soft limits,''
  arXiv:1406.1462 [hep-th].


  \bibitem{lowsym} 
  V.~Lysov, S.~Pasterski and A.~Strominger,
 ``Low's Subleading Soft Theorem as a Symmetry of QED,''
  Phys.\ Rev.\ Lett.\  {\bf 113}, 111601 (2014)
  [arXiv:1407.3814 [hep-th]]


\bibitem{am}
A.~Ashtekar and A.~Magnon-Ashtekar
``On the symplectic structure of general relativity''
Commun.\ Math.\ Phys.\  {\bf 86}, 55 (1982).

\bibitem{aaunpublished}
A.~Ashtekar, unpublished notes.
\bibitem{frolov} 
  V.~P.~Frolov,
  ``Null Surface Quantization and Quantum Field Theory in Asymptotically Flat Space-Time,''
  Fortsch.\ Phys.\  {\bf 26}, 455 (1978).

  \bibitem{dp1} 
  C.~Dappiaggi, V.~Moretti and N.~Pinamonti,
  ``Rigorous steps towards holography in asymptotically flat spacetimes,''
  Rev.\ Math.\ Phys.\  {\bf 18}, 349 (2006)
  [gr-qc/0506069]
  

  
  \end{thebibliography}
\end{document}